\documentclass[twocolumn,aps,prc,superscriptaddress,showpacs]{revtex4}
\usepackage{amssymb}
\usepackage{amsmath}
\usepackage{graphicx}
\usepackage[normalem]{ulem}
\usepackage{multirow}
\usepackage{appendix}
\usepackage{CJK}
\usepackage[usenames]{color}
\usepackage{bm}
\makeatletter % `@' now normal "letter"
\@addtoreset{equation}{section}
\makeatother  % `@' is restored as "non-letter"

\begin{document}
\begin{CJK*}{GBK}{song}

\title{
Directed flow in an extended multiphase transport model
}

\author{Chong-Qiang Guo}
\affiliation{Shanghai Institute of Applied Physics, Chinese Academy of Sciences, Shanghai 201800, China}
\affiliation{University of Chinese Academy of Sciences, Beijing 100049, China}
\author{He Liu}
\affiliation{Shanghai Institute of Applied Physics, Chinese Academy of Sciences, Shanghai 201800, China}
\affiliation{University of Chinese Academy of Sciences, Beijing 100049, China}
\author{Jun Xu\footnote{Corresponding author: xujun@sinap.ac.cn}}
\affiliation{Shanghai Institute of Applied Physics, Chinese Academy of Sciences, Shanghai 201800, China}

\date{\today}

\begin{abstract}

We have studied the rapidity-odd directed flow in $^{197}$Au+$^{197}$Au collisions in the beam energy range from $\sqrt{s_{NN}}$ = 7.7 to 39 GeV within the framework of an extended multiphase transport model with both partonic and hadronic mean-field potentials incorporated. Effects of the partonic scatterings, mean-field potentials, hadronization, and hadronic evolution on the directed flow are investigated, and it is found that the final directed flow is mostly sensitive to the partonic scatterings and the hadronization mechanism. Our study shows that a negative slope of the proton directed flow does not necessarily need the equation of state with a first-order phase transition.
\end{abstract}


\maketitle

\section{Introduction}

Understanding the properties of the quark-gluon plasma as well as the hadron-quark phase transition is one of the main purposes of relativistic heavy-ion collision experiments. The directed flow $v_1$, especially its rapidity-odd component to be discussed in the present study, is an important probe characterizing the dynamics in these collisions. Both hydrodynamic~\cite{hydro} and transport model~\cite{transport} studies have shown that the baryon $v_1(y)$ in the midrapidity region ($y\sim0$) is sensitive to the equation of state (EoS) of the produced matter. With the increasing collision energies, these calculations predicted that the slope of $v_1(y)$ near midrapidity region changes from positive to negative~\cite{antiflow,thirdflow,wiggle}, and argued that this is the result of a soft EoS due to the first-order hadron-quark phase transition. The recent directed flow results from RHIC beam energy scan (BES) program~\cite{RHIC1,RHIC2} seem to support this argument, where the slope of the proton directed flow changes sign from positive to negative between $\sqrt{s_{NN}}$ = 7.7 and 11.5 GeV, while the slope of the net proton directed flow changes sign twice between $\sqrt{s_{NN}}$ = 11.5 and 39 GeV, and has a minimum between $\sqrt{s_{NN}}$ = 11.5 and 19.6 GeV. Considerable efforts have been devoted to this topic with various approaches, e.g., a hybrid approach with the fluid dynamics model for the partonic phase and the ultra-relativistic quantum molecular dynamics model for the hadronic phase~\cite{Ste14}, the parton-hadron-string-dynamics model and a 3-fluid hydrodynamics approach~\cite{Kon14}, and a pure hadronic transport approach but with the collision term modified to mimic the softness of the EoS~\cite{Nar16}. However, none of them have described the experimental data of the directed flow at various collision energies very satisfactorily.

Obviously, so far the studies have shown that the relation between the $v_1(y)$ slope and the EoS is not as simple as expected, and the former is also sensitive to other factors, which needs to be further investigated in detail before a definite conclusion can be drawn. For this purpose, we investigate the directed flow in relativistic heavy-ion collisions based on the framework of an extended multiphase transport (AMPT) model~\cite{AMPT_MF}, which was developed from the original AMPT model~\cite{AMPT} by incorporating the partonic mean-field potential based on the 3-flavor Nambu-Jona-Lasinio (NJL) model~\cite{Son12,Xu14} and the hadronic mean-field potential based the relativistic mean-field model and the chiral effective field theory~\cite{Xu12}. This model has been used to study the elliptic flow splitting between particles and their antiparticles due to their different mean-field potentials~\cite{Xu14,AMPT_MF}, and thus it is of great interest to see how the mean-field potentials affect their directed flows. Furthermore, this model provides the possibility to investigate in detail the effects from the partonic scatterings, hadronization, and the hadronic interaction on the directed flow. We found that the mean-field potential, which is related to the EoS through the energy density functional, has only moderate effect on the directed flow, while the partonic scatterings and the hadronization mechanism dominate the final directed flow.

The rest of the paper is organized as follows. Section \ref{sec:2} provides a brief description of the structure of the extended AMPT model. The detailed analysis and discussions of the directed flow results are given in Sec.~\ref{sec:3}. A summary and final remark is given in Sec.~\ref{sec:4}.

\section{\label{sec:2} An extended AMPT model}

The initial momentum distribution of partons in the extended AMPT model is generated by melting hadrons from the heavy-ion jet interaction generator (HIJING) model~\cite{hijing}, while the spatial distribution of these partons are modified since they are expected to be more expanded in the beam direction at the RHIC-BES energies, compared to the treatment in the original AMPT model for ultra-relativistic heavy-ion collisions. Similar to Ref.~\cite{lif2017}, we sample the longitudinal coordinate of initial partons uniformly within $(-lm_N/\sqrt{s_{NN}}, lm_N/\sqrt{s_{NN}})$, where $l=14$ fm is approximately the diameter of the Au nucleus, and $m_N=0.938$ GeV is the nucleon mass. A more realistic treatment of the finite thickness effect can be found in Ref.~\cite{Lin17}. The evolution of the partonic phase is described by a 3-flavor NJL transport model, with a Lagrangian given by~\cite{Son12,Xu14}
\begin{eqnarray}
\mathcal{L_{NJL}} &=& \bar{q}(i\not{\partial}-M)q+\frac{G_S}{2}\sum_{a=0}^{8}\bigg[(\bar{q}\lambda^aq)^2+(\bar{q}i\gamma_5\lambda^aq)^2\bigg]\nonumber\\
&-&\frac{G_V}{2}\sum_{a=0}^{8}\bigg[(\bar{q}\gamma_\mu\lambda^aq)^2+(\bar{q}\gamma_\mu\gamma_5\lambda^aq)^2\bigg]\nonumber\\
&-&K\bigg[{\rm det}_f\bigg(\bar{q}(1+\gamma_5)q\bigg)+{\rm
det}_f\bigg(\bar{q}(1-\gamma_5)q\bigg)\bigg],
\label{Lagrangian}
\end{eqnarray}
where $q=(u, d, s)^T$ is the quark fields, $M={\rm diag}(m_u, m_d, m_s)$ is the current quark mass matrix in flavor space, and $\lambda^{a}$ is the Gell-Mann matrices in $SU(3)$ flavor space with $\lambda^0=\sqrt{2/3}I$. $G_S$ and $G_V$ are the strength of the scalar and vector coupling, respectively.
The last term in Eq.~(\ref{Lagrangian}), with det$_f$ denoting the determinant in the flavor space, is the Kobayashi-Maskawa-t'Hooft (KMT) interaction~\cite{Hoo76}. In the present study, we employ the parameters of the current quark mass $m_u=m_d=3.6$ MeV and $m_s=87$ MeV, and the values of coupling constants $G_S$ and $K$ from $G_S\Lambda^2=3.6$ and $K\Lambda^{5}=8.9$, with the cutoff value in the momentum integral $\Lambda=750$ MeV given in Refs.~\cite{Lut92,Bra13}. As in the previous studies, we define $R_V=G_V/G_S$ as the relative strength of the vector coupling. From the mean-field approximation and some algebras based on the finite-temperature field theory, we can extract the single-quark Hamiltonian and the thermodynamic properties of quark matter (see more details in \ref{app}). The scalar mean-field potential enlarges the in-medium quark mass through the quark condensate, and it is the same for quarks and antiquarks and attractive in the non-relativistic reduction. The time component of the vector potential for positive $G_V$ value is repulsive for quarks and attractive for antiquarks, while its space component has the opposite effect. A constant and isotropic cross section of 3 mb is used for the parton elastic scattering process, and it is the same for all kinds of partons. The test-particle method~\cite{Won82,Ber88} is used to calculate the average phase-space distribution function and thus the mean-field potential, from averaging over parallel events at the same impact parameter. A mix-event spatial coalescence algorithm is used to describe the hadronization, which allows quarks and antiquarks in one event to coalesce with nearest quarks or antiquarks from all parallel events at the end of the partonic phase, when the central energy density becomes lower than about $0.8$ GeV/fm$^{3}$. Combinations from three nearest quarks (antiquarks) are chosen to form baryons (antibaryons), before nearest quark and antiquark pairs from the rest partons are chosen to form mesons, and the species of formed hadrons are determined by both the flavors and the invariant mass of their valence quarks and/or antiquarks. The hadronization scheme produces not only ground-state hadrons but also excited-state resonances with a finite mass width. For more details about the hadron production in the AMPT model, we refer the reader to Ref.~\cite{AMPT}. An additional formation time of about 0.7 fm/c is used for all hadrons. After the formation of initial hadrons, a relativistic transport model (ART)~\cite{art} with various elastic, inelastic, and decay channels is used to describe the evolution of the hadronic phase, where the mean-field potentials for hadrons are further implemented~\cite{Xu12}. The mean-field potentials for nucleons and antinucleons are from the relativistic mean-field theory based on G-parity invariance~\cite{hadronicMF}. The mean-field potentials for kaons and antikaons are from the chiral effective Lagrangian~\cite{Li97}. The $s$-wave mean-field potentials for pions are from the chiral perturbation theory up to the two-loop order~\cite{Kai01}. For more details about this model, we refer the reader to Ref.~\cite{AMPT_MF}.

\section{\label{sec:3} Results and analysis}

In the present study, we employ the extended multiphase transport model to investigate the directed flow in midcentral ($c = 10 - 40\%$) $^{197}$Au+$^{197}$Au collisions at the typical RHIC-BES energies, i.e., $\sqrt{s_{NN}}$ = 7.7, 11.5, 19.6, 27, and 39 GeV. With the total inelastic cross section $\sigma_{in}$ of about 705 fm$^2$ for Au+Au collisions, the corresponding impact parameters are about $\text{b}=4.7 - 9.5$ fm, from the empirical relation $c=\pi b^2/\sigma_{in}$~\cite{Bro02}. From the analysis in \ref{app}, the partonic phase doesn't pass through the spinodal region, which corresponds to the softening of the EOS, at the above collision energies from the parameterization of the NJL model. From the description of the hadronization process in Sec.\ref{sec:2}, there is no hadron-quark mixed phase, and thus no softness of the EoS during the phase transition, either. In the following analysis, the directed flow is calculated by averaging the azimuthal angle $\phi$ of particle momenta with respect to the theoretical reaction plane, i.e., $v_1=\langle \cos(\phi) \rangle$.

\subsection{Directed flow in the partonic phase}

\begin{figure}[h]
\includegraphics[scale=0.3]{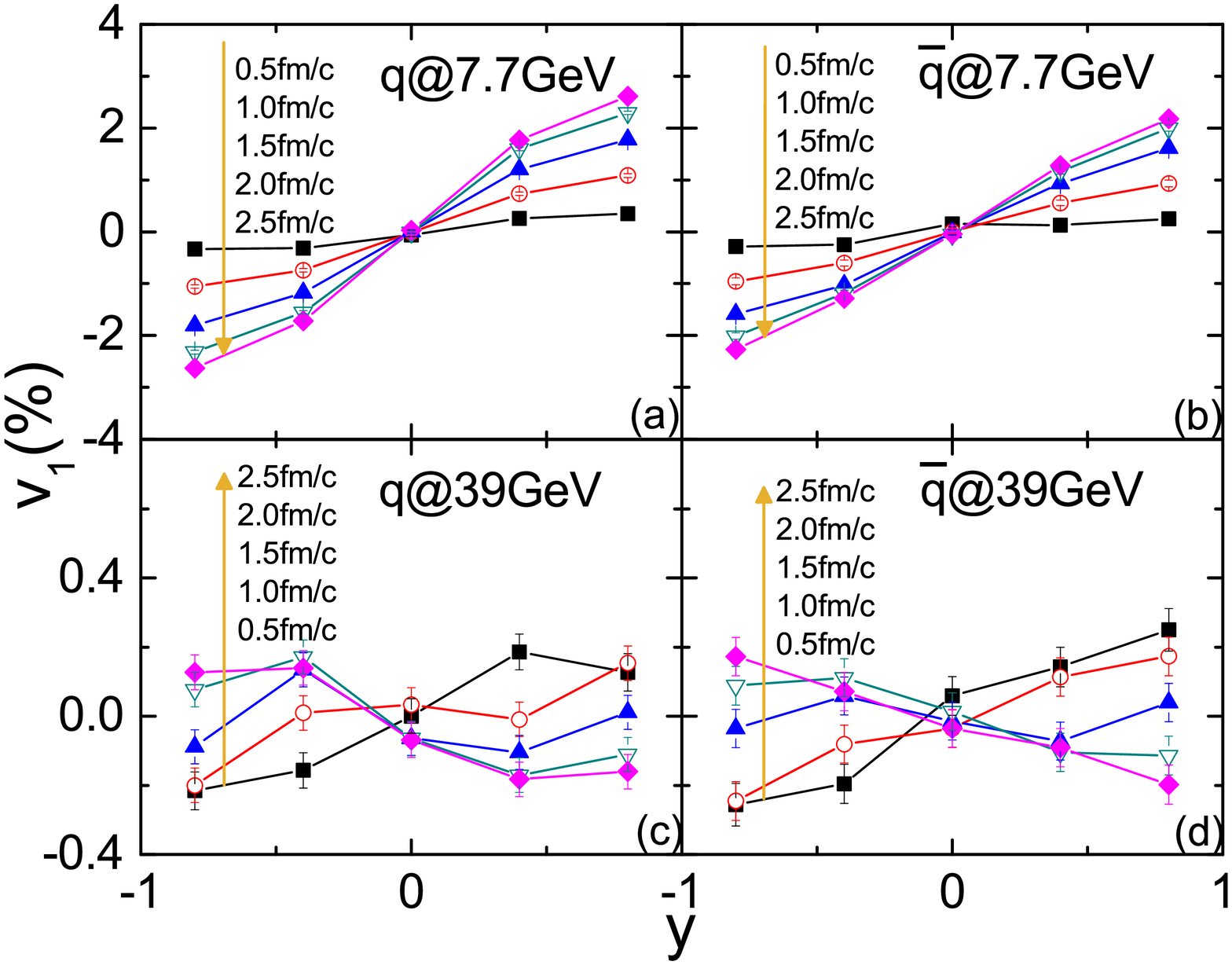}
\caption{(Color online) Directed flow $v_1$ of quarks [(a), (c)] and antiquarks [(b), (d)] versus rapidity $y$ at different times in midcentral Au+Au collisions at $\sqrt{s_{NN}}$ = 7.7 [(a), (b)] and 39 GeV [(c), (d)]. Note the different scales for $v_1$ at 7.7 and 39 GeV.}
\label{F1}
\end{figure}

To begin with the discussion on the directed flow in the partonic phase, we first show the directed flow of quarks and antiquarks in midcentral Au+Au collisions at $\sqrt{s_{NN}}$ = 7.7 and 39 GeV at different time steps in Fig.~\ref{F1}, with the parton scattering cross section 3 mb together the scalar potential but without the vector potential. In the early stage of $t=0.5$ fm/c, the $v_1$ slopes of quarks and antiquarks at midrapidities at both collision energies are small positive values. As the system evolves, the slope of the directed flow at $\sqrt{s_{NN}}$ = 39 GeV changes sign at around 1.5 fm/c and then becomes saturated, while the slope at $\sqrt{s_{NN}}$ = 7.7 GeV keeps growing to a maximum positive value. The behavior of the directed flow at 39 GeV is similar to that observed in Ref.~\cite{guoepja}, where the $v_1$ slope at midrapidities keeps going negative and then saturates in the later stage at 39 GeV due to repulsive partonic scatterings in the absence of the mean-field potential. This behavior is due to the transfer of partons among different rapidity regions~\cite{guoepja}, i.e., partons contributing to the positive flow are scattered to large rapidities and only those contributing to the negative flow stay at midrapidities. At 7.7 GeV, the scatterings are not strong enough for the parton transfer among different rapidity regions due to the lower parton density/pressure, and the slope of the directed flow keeps going positive. Here we have already observed that the directed flow of partons changes sign with the increasing collision energy without a first-order phase transition. The final parton directed flow is not expected to change by much if we vary the criterion for the end of the partonic evolution, since it is mostly saturated around $t=2.5$ fm/c.

\begin{figure}[h]
\includegraphics[scale=0.3]{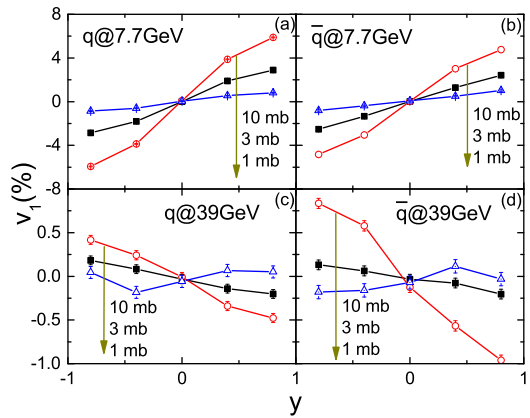}
\caption{(Color online) Directed flow $v_1$ of freeze-out quarks [(a), (c)] and antiquarks [(b), (d)] versus rapidity $y$ from the parton scattering cross sections of $\sigma$ = 1, 3, and 10 mb in midcentral Au+Au collisions at $\sqrt{s_{NN}}$ = 7.7 [(a), (b)] and 39 GeV [(c), (d)]. Note the different scales for $v_1$ at 7.7 and 39 GeV.}
\label{F4}
\end{figure}

From our argument above, the partonic scatterings are important in determining the $v_1$ slope of freeze-out partons. Figure~\ref{F4} displays the directed flow of final quarks and antiquarks from the parton scattering cross sections of 1, 3, and 10 mb together with the scalar potential but in the absence of the vector potential. A larger parton scattering cross section is expected to enhance the interaction in both the early stage and the later stage, depending on the parton number density. It is seen that the slope of the directed flow becomes more positive at 7.7 GeV but more negative at 39 GeV from a larger parton scattering cross section. At 39 GeV, it is interesting to see that a small cross section of 1 mb leads to a positive slope of $v_1$, while a larger cross section leads to a negative slope. The partonic scatterings thus dominate the directed flow of final partons, or even change the slope sign especially at higher collision energies. In addition, the stronger interaction from partonic scatterings reveals more explicitly the difference in the initial phase-space distributions between quarks and antiquarks, and thus their final directed flows.

\begin{figure}[h]
\includegraphics[scale=0.3]{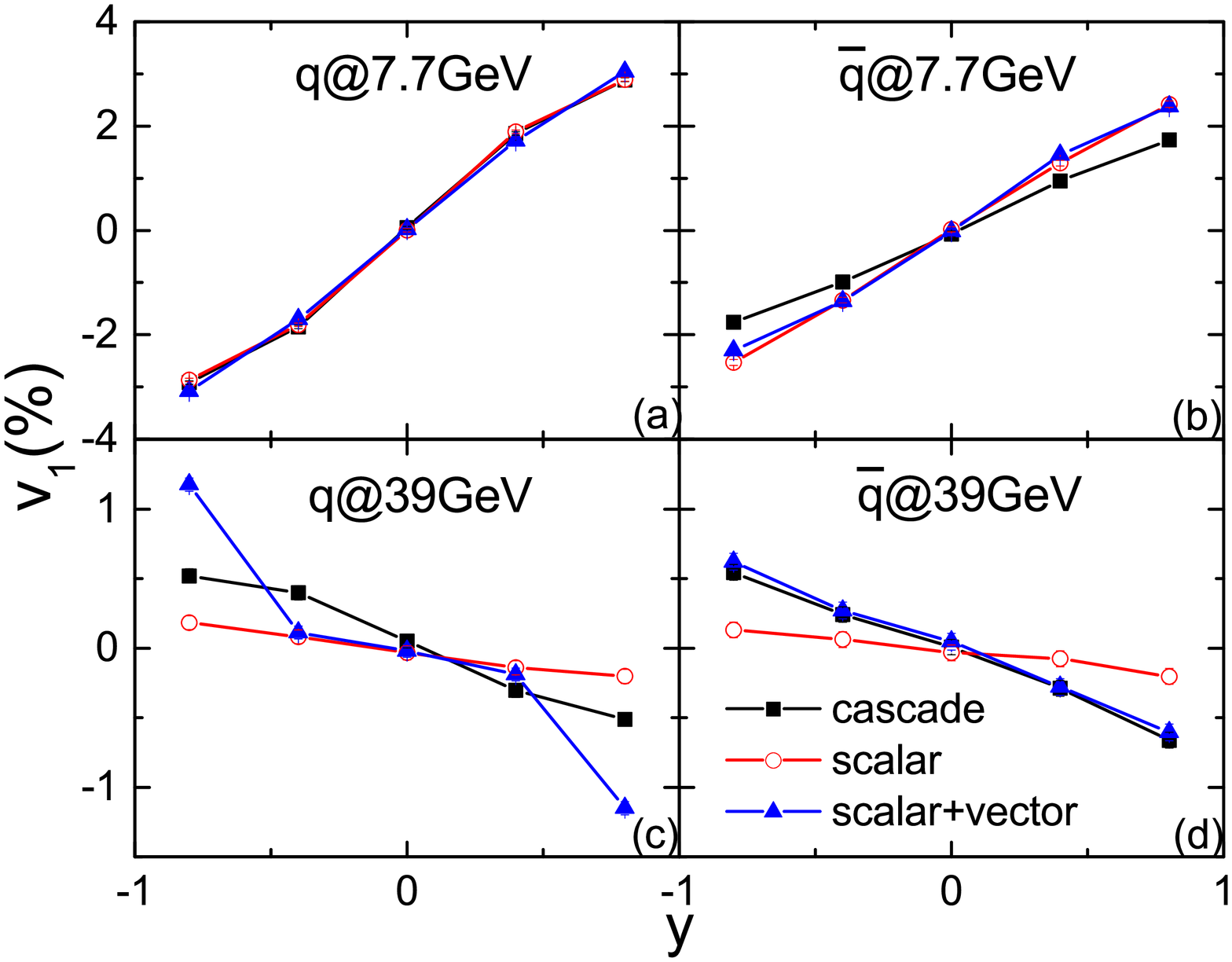}
\caption{(Color online) Directed flow $v_1$ of freeze-out quarks [(a), (c)] and antiquarks [(b), (d)] versus rapidity $y$ from only partonic scatterings (cascade), partonic scatterings together with the scalar potential (scalar), and partonic scatterings together with both scalar and vector potentials (scalar+vector) in midcentral Au+Au collisions at $\sqrt{s_{NN}}$ = 7.7 [(a), (c)] and 39 GeV [(b), (d)]. Note the different scales for $v_1$ at 7.7 and 39 GeV.}
\label{F3}
\end{figure}

In order to understand the effects of the mean-field potential on the parton directed flow, we display in Fig.~\ref{F3} the directed flow of final quarks and antiquarks with only partonic scatterings, partonic scatterings together with the scalar potential, and partonic scatterings together with both scalar and vector potentials with $R_V = 1.1$ at $\sqrt{s_{NN}}$ = 7.7 GeV and 39 GeV. It is found that neither the scalar potential nor the vector potential changes the slope sign of parton $v_1$, and this shows that the mean-field potential has only moderate effects on the parton directed flow. It is also seen that the scalar and the vector potential have different effects on the directed flows of quarks and antiquarks. The attractive scalar potential leads to less negative slope of the directed flow for both quarks and antiquarks at 39 GeV. At 7.7 GeV, it has small effects on the directed flow of quarks, but somehow enhances the directed flow of antiquarks, likely due to the enhancement of the scatterings. It is seen that the vector potential has a larger effect on the directed flow at 39 GeV than at 7.7 GeV, and the effect can be seen at large rapidities for quarks and in the whole rapidity region for antiquarks. Note that the scalar potential and the time component of the vector potential related to the quark number density are stronger in the early stage but weaker in the later stage, while the space component of the vector potential related to the quark flux density is expected to be weaker in the early stage but stronger in the later stage. The results displayed here show that there are interplays between effects from the mean-field potentials and partonic scatterings, and the effect of the mean-field potential on the directed flow is different from that on the elliptic flow~\cite{Xu14,AMPT_MF} due to their different dynamic mechanisms.

\subsection{Directed flow in the hadronic phase}

\begin{figure}[h]
\includegraphics[scale=0.3]{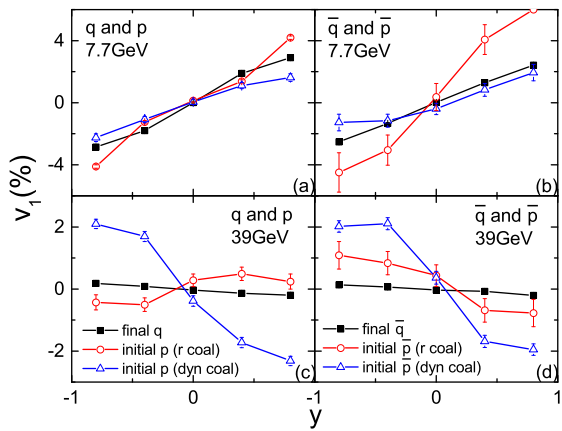}
\caption{(Color online) Directed flow $v_1$ of final partons and initial protons as well as initial antiprotons from the mix-event spatial coalescence approach in the extended AMPT model (r coal) and the dynamical coalescence approach (dyn coal) versus rapidity $y$ in midcentral Au+Au collisions at $\sqrt{s_{NN}}$ = 7.7 [(a), (b)] and 39 GeV [(c), (d)]. Note the different scales for $v_1$ at 7.7 and 39 GeV.}
\label{F5}
\end{figure}

The directed flow of final partons discussed in previous subsections will be further modified in the hadronization process and the hadronic evolution, and this has been investigated in our previous study~\cite{guoepja} in the absence of the mean-field potential. Based on the framework of the extended AMPT model, we compared the directed flows of initial protons and antiprotons from the mix-event spatial coalescence approach as in the extended AMPT model and a dynamical coalescence approach~\cite{dynamic,splitting} in Fig.~\ref{F5}. In the dynamical coalescence approach, the probability to form a hadron is proportional to the quark Wigner function of that hadron, and  partons that are close in phase space, i.e., both coordinate and momentum space, have a larger probability to form hadrons. The dynamical coalescence approach has been successfully used to explain reasonable well spectra and elliptic flows of hadrons~\cite{dynamic}. In the mix-event spatial coalescence approach as employed in the extended AMPT model, hadrons are formed by nearest combinations of partons in coordinate space in parallel events, and all the partons are force to be used up after hadronization. The difference of the directed flows for protons and antiprotons from the two hadronization approaches is observed, especially for protons at 39 GeV where different slope signs are obtained. As discussed in detail in Ref.~\cite{guoepja}, a pure coalescence in momentum space always keeps the slope sign of the directed flow from final partons to initial hadrons, while there are competition effect from the coalescence in coordinate space, and this depends on the phase-space distribution in the freeze-out stage of the partonic evolution. With the Gaussian width of the Wigner function fitted by the root-mean-square radius of the hadron, the dynamical coalescence approach is expected to describe more properly the competition effect of the coalescence in both the coordinate and momentum space. Our finding that the hadron directed flow is sensitive to the hadronization scenario is consistent with that in Ref.~\cite{Ste14}. So far the dynamical coalescence approach is limited to the study of the directed flow of initial hadrons without further hadronic evolution. Incorporations of a more realistic hadronization criterion together with a proper hadronization scenario in the transport model need further investigations to give unambiguous results.

\begin{figure}[h]
\includegraphics[scale=0.3]{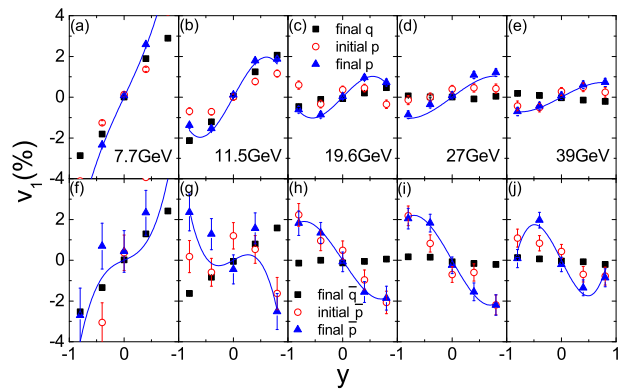}
\caption{(Color online) Directed flow $v_1$ of final partons as well as initial and final protons and antiprotons versus rapidity $y$ in midcentral Au+Au collisions at $\sqrt{s_{NN}}$ = 7.7, 11.5, 19.6, 27, and 39 GeV.}
\label{F6}
\end{figure}

The directed flows of final partons as well as the initial and final protons and antiprotons from the mix-event spatial coalescence and the hadronic evolution in midcentral Au+Au collisions at $\sqrt{s_{NN}}$ = 7.7, 11.5, 19.6, 27, and 39 GeV are compared in Fig.~\ref{F6}, where the solid lines are from a cubic fit of the rapidity dependence of the final directed flow, i.e., $v_1(y) = F_1 y + F_3y^3$. From the spatial coalescence approach, the $v_1$ slopes of initial protons are positive at all energies, although the directed flows of final quarks have negative slopes at higher collision energies. On the other hand, the hadronic evolution with the hadronic mean-field potentials properly incorporated only slightly enhances the magnitude the directed flow, no matter whether its slope is positive or negative.

%\subsection{slope of the $v_1$ at midrapidity}

\begin{figure}[h]
\includegraphics[scale=0.3]{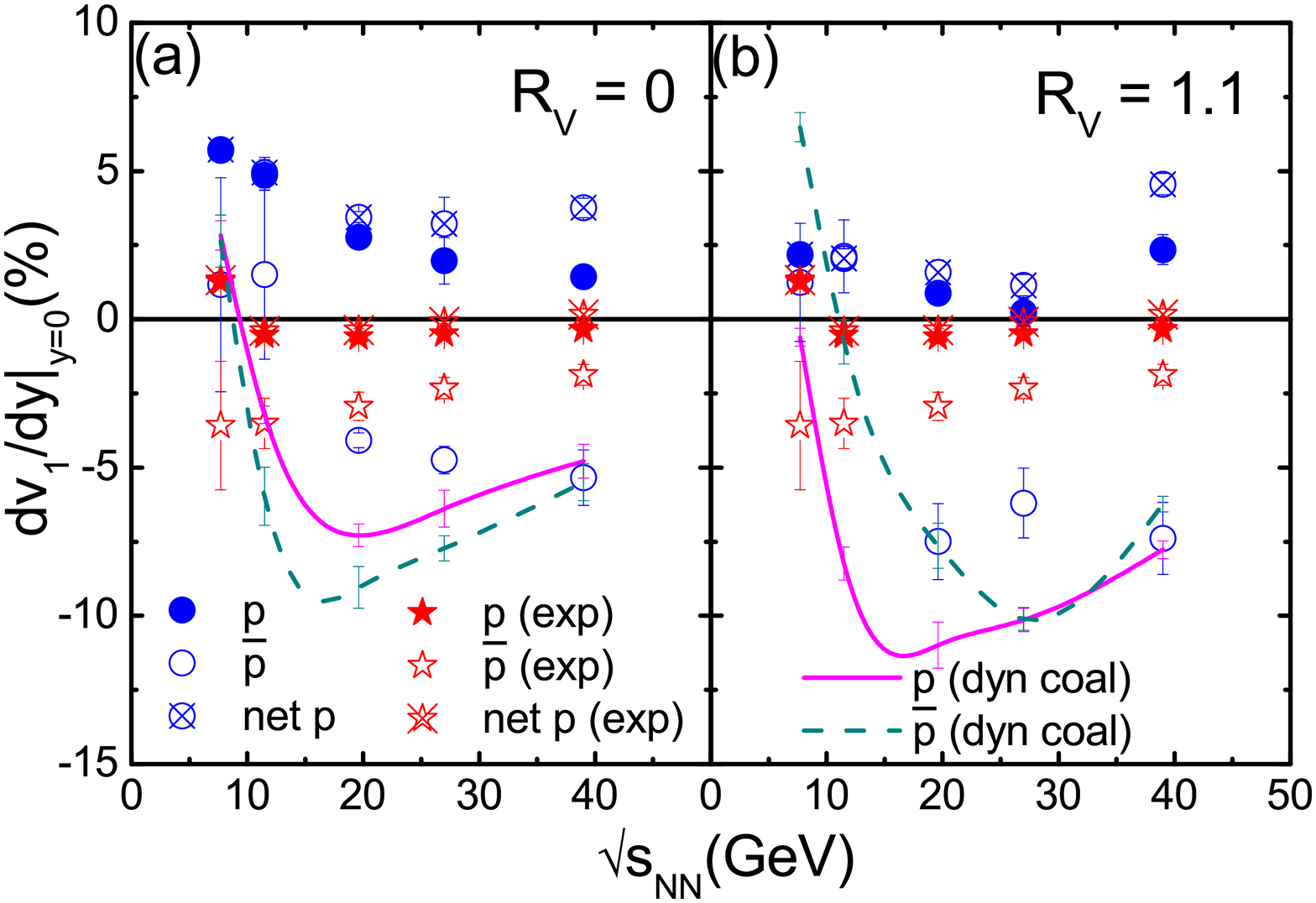}
\caption{(Color online) Directed flow slope $dv_1/dy|_{y=0}$ of protons, antiprotons, and net protons versus collisions energy in midcentral Au+Au collisions with $R_V = 0$ (a) and $R_V = 1.1$ (b). Results of star symbols are experimental data measured by the STAR Collaboration~\cite{RHIC1}, and the lines are the directed flow slope from the dynamical coalescence approach.}
\label{F7}
\end{figure}

We summarize the energy dependence of the directed flow slopes at midrapidities for protons and antiprotons in Fig.~\ref{F7}, and the experimental results from the STAR Collaboration~\cite{RHIC1} are also plotted in the same figure. From the spatial coalescence at higher collision energies, the proton directed flows always have a positive slope, although the slope of freeze-out quark $v_1$ is negative at higher collision energies, as can be seen from Figs.~\ref{F5} and \ref{F6}. The slope of the antiproton directed flow is positive at lower collision energies but rather negative at higher collision energies. From a dynamical coalescence approach but in the absence of the hadronic evolution, the directed flows of both protons and antiprotons have a positive slope at lower collision energies but a negative slope at higher collision energies, reflecting the $v_1$ slope of quarks and antiquarks in the final stage of the partonic phase. The hadronic evolution after the hadronization described by the dynamical coalescence approach, if properly incorporated, is expected to modify slightly the magnitude but not change the slope sign of the directed flow. Results with the vector potential ($R_V=1.1$) show a similar behavior from the spatial coalescence approach, while it changes the relative slope of proton and antiproton $v_1$ from the dynamical coalescence approach. In order to obtain a negative $v_1$ slope of protons or net protons, the coalescence in momentum space, such as that in the dynamical coalescence approach, should be taken into account, while the EoS with a first-order phase transition is not necessarily needed.

\section{\label{sec:4} Summary}

We have studied in detail the effects of partonic scatterings as well as the mean-filed potential, hadronization, and hadronic evolution on the directed flow in relativistic heavy-ion collisions within the framework of the extended multiphase transport model. As the system evolves, a non-monotonic behavior of the directed flow is observed at higher collision energies, as a result the later dynamics which changes the slope sign of the directed flow from positive to negative at midrapidity region. The final directed flow of partons, particularly the sign of its slope at midrapidities, is dominated by the partonic scatterings especially at higher collision energies. The mean-field potentials in the partonic phase, which can be attractive or repulsive, have only moderate effects on the parton directed flow. The hadron directed flow is significantly affected by the hadronization mechanism, while it is only slightly affected by the hadronic evolution with the mean-field potentials for hadrons incorporated.

Based on this study, we found that a large parton scattering cross section and a coalescence approach in momentum space may result in a negative slope of the final directed flow. Although the EoS with a first-order phase transition may lead to a negative slope of the directed flow, the present study shows that the negative slope of the directed flow does not necessarily need the softness of the EoS. Before drawing a conclusion on the EoS and the order of the hadron-quark phase transition from the experimental data, deeper understandings on various stages in relativistic heavy-ion collisions are needed.

\begin{acknowledgements}
We thank Chen Zhong for maintaining the high-quality performance of the computer facility.
This work was supported by the Major State Basic Research Development Program (973 Program) of China under Contract Nos. 2015CB856904 and 2014CB845401, the National Natural Science Foundation of China under Grant Nos. 11475243 and 11421505, and the Shanghai Key Laboratory of Particle Physics and Cosmology under Grant No. 15DZ2272100.
\end{acknowledgements}

\appendices

\renewcommand\thesection{APPENDIX~\Alph{section}}
\renewcommand\theequation{\Alph{section}.\arabic{equation}}
\section{Formulism of the NJL model}
\label{app}

The single-particle Hamiltonian for quarks (antiquarks) with flavor $i$ ($i=u, d, s$) from the NJL Lagrangian [Eq.~(\ref{Lagrangian})] is written as~\footnote{Here only the flavor-singlet state of the vector interaction is considered. In this way, quarks and antiquarks are affected by the vector mean-field potential generated by partons of different flavors.}
\begin{equation}\label{h}
H_i = \sqrt{M_i^2+{p^*}^2} \pm \frac{2}{3}G_V \rho^0,
\end{equation}
where $M_i$ is the quark constituent mass, $\vec{p^*} = \vec{p} \mp \frac{2}{3}G_V \vec{\rho}$ is the real momentum of the particle with $\vec{\rho}$ being the space component of the vector density, and $\rho^0$ is the time component of the vector density. The upper (lower) sign in the above equations are for quarks (antiquarks). The quark constituent mass $M_i$ is given by the gap equation as
\begin{eqnarray}
M_i &=&
m_i-2G_S\sigma_i+2K\sigma_j\sigma_k,
\label{mi}
\end{eqnarray}
where $\sigma_i$ is the quark condensate expressed as
\begin{eqnarray}
\sigma_i&=&-2N_c\int_0^\Lambda\frac{d^3p}{(2\pi)^3}\frac{M_i}{E_i}(1-f_i-\bar{f}_i),
\label{sigma}
\end{eqnarray}
where the factor $2N_c$ represents the spin and color degeneracy, $E_i=\sqrt{{p^*}^2 +M_i^2}$ is the quark energy, and $f_i$ and $\bar{f}_i$ are respectively the phase-space distributions of quarks and antiquarks. They can be obtained by counting parton numbers in the local phase-space cell through the test-particle method, and in the thermodynamic limit they can be expressed as the Fermi-Dirac distributions, i.e.,
\begin{eqnarray}
f_i &=& \frac{1}{1+e^{\beta(E_i-\tilde{\mu}_i)}},
\\
\bar{f_i} &=& \frac{1}{1+e^{\beta(E_i+\tilde{\mu}_i)}},
\end{eqnarray}
where $\beta=1/T$ represents the temperature, and the effective chemical potential $\tilde{\mu}_i$ is defined as
\begin{eqnarray}
\tilde{\mu}_i&=&\mu_i-\frac{2}{3}G_V \rho^0.
\label{mui}
\end{eqnarray}
The 4-component net quark number density of the flavor $i$ can be calculated from $f_i$ and $\bar{f}_i$ via
\begin{eqnarray}
\rho^\nu_i=2N_c\int^\Lambda_0(f_i-\bar{f_i})\frac{p_i^\nu}{E_i}\frac{d^3p}{(2\pi)^3},
\label{rho}
\end{eqnarray}
with $\rho^\nu = \rho^\nu_u + \rho^\nu_d + \rho^\nu_s$ being the total number density. The time component ($\nu=0$) of the above density is the net quark number density, while the space component ($\nu=1,2,3$) is the net quark flux density. As seen from Eq.~(\ref{mi}), the scalar mean-field potential is generated from the quark condensate. The space and time components of the vector mean-field potential are $\mp \frac{2}{3} G_V \vec{\rho}$ and $\pm \frac{2}{3} G_V \rho^0$ terms in Eq.~(\ref{h}), respectively.

From the mean-field approximation and some algebras based on the finite-temperature field theory, the thermodynamic potential $\Omega_{NJL}$ of static quark matter at finite temperature and quark chemical potential can be expressed as
\begin{eqnarray}
\Omega_{NJL} &=& -2N_c\sum_{i=u,d,s}\int_0^\Lambda\frac{d^3p}{(2\pi)^3}
[E_i+T\ln(1+e^{-\beta(E_i-\tilde{\mu}_i)})
\notag\\
&+&T\ln(1+e^{-\beta(E_i+\tilde{\mu}_i)})]+G_S(\sigma_u^2+\sigma_d^2+\sigma_s^2)
\notag\\
&-&4K\sigma_u\sigma_d\sigma_s-\frac{1}{3}G_V{\rho^0}^2.
\label{omega}
\end{eqnarray}
The energy density $\varepsilon_{NJL}$ from the NJL model can
be written as
\begin{eqnarray}
\varepsilon_{NJL} &=& -2N_c\sum_{i=u,d,s}\int_0^\Lambda\frac{d^3p}{(2\pi)^3}
E_i(1-f_i-\bar{f}_i)
\notag\\
&-& \sum_{i=u,d,s}(\tilde{\mu}_i-\mu_i)\rho^0+G_S(\sigma_u^2+\sigma_d^2+\sigma_s^2)
\notag\\
&-& 4K\sigma_u\sigma_d\sigma_s-\frac{1}{3}G_V{\rho^0}^2-\varepsilon_0,
\label{epsilon}
\end{eqnarray}
where $\varepsilon_0$ is introduced to ensure $\varepsilon_{NJL} = 0$ in vacuum.

\begin{figure}[h]
\includegraphics[scale=0.3]{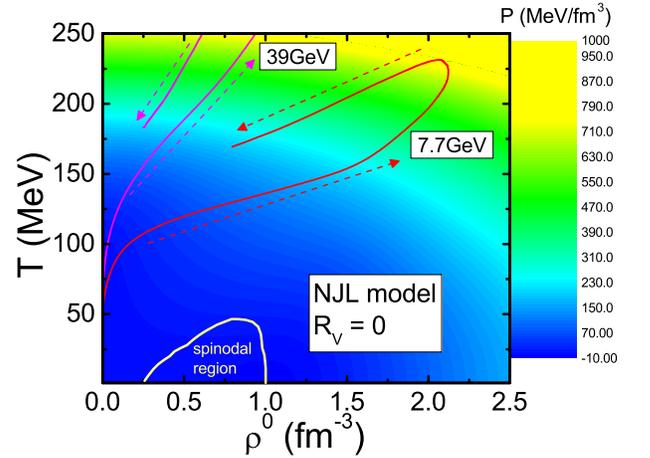}
\caption{(Color online) Equation of state in the [number density ($\rho^0$), temperature ($T$)] plane of the partonic phase from the NJL model with $R_V = 0$, and the corresponding trajectories of the central region of the partonic phase in midcentral Au+Au collisions at $\sqrt{s_{NN}}$ = 7.7 and 39 GeV based on the extended AMPT model. The spinodal region with $(\partial P/\partial \rho^0)_T<0$ is also approximately marked.}
\label{F8}
\end{figure}

For a given net quark number density $\rho^0$ and temperature $T$, the energy density $\epsilon_{NJL}$ of quark matter can be obtained from Eq.~(\ref{epsilon}) by solving self-consistently Eqs.~(\ref{mi})$-$(\ref{rho}). In the partonic phase described by the NJL transport model, by assuming that the central region is static and in thermal equilibrium, we can calculate the temperature $T$ inversely from the energy density $\epsilon_{NJL}$ and the net quark number density $\rho^0$. In this way, we can obtain a trajectory in the $(\rho^0, T)$ plane for the central region of the partonic phase in relativistic heavy-ion collisions, and this is shown in Fig.~\ref{F8} for midcentral Au+Au collisions at $\sqrt{s_{NN}}$ = 7.7 and 39 GeV, where the pressure $P = -\Omega_{NJL}$ in the $(\rho^0, T)$ plane is also displayed. Due to the different NJL parameterizations and collision energies in the present study compared to that in Ref.~\cite{lif2017}, neither of the trajectories enters the spinodal region, i.e., $(\partial P/\partial \rho^0)_T<0$, as marked approximately in Fig.~\ref{F8}. This means that the central region in the partonic phase doesn't pass through a first-order liquid-gas phase transition in this study. The EoS from $R_V=1.1$ is even stiffer, and the spinodal region disappears~\cite{lif2017}.

\end{CJK*}
\end{document}